\documentclass[10pt,preprint,a4paper]{aastex}
\usepackage{amsmath}                
\usepackage{amsfonts}               
\usepackage{amssymb}                
\usepackage{epsfig}                 
\usepackage{float}
\usepackage{graphicx}

\begin{document}

\def \kms{\rm{km}$\rm{s}^{-1}$}
\def \kev{\rm{keV}}
\def \eV{~\rm{eV}}
\def \cmcub {$\rm{cm}^{-3}$}
\def \msyr{$M_{\odot}~yr^{-1}$}
\def \cm{~\rm{cm}}
\def \s{~\rm{s}}
\def \km{~\rm{km}}
\def \kms{$~\rm{km}~{\rm s}^{-1}$}
\def \gm{\rm{gm}}
\def \K{~\rm{K}}
\def \g{~\rm{g}}
\def \G{~\rm{G}}
\def \AU{~\rm{AU}}
\def \erg{~\rm{erg}}
\def \foe{~\rm{foe}}
\def \yrs{~\rm{yrs}}
\def \yr{~\rm{yr}}
\def \pc{~\rm{pc}}
\def \kpc{~\rm{kpc}}
\def \etc{$\eta$~Car}
\def \days{~\rm{day}}
\def \Jy{~\rm{Jy}}
\def \mum{~\rm{\mu m}}
\def \keV{~\rm{keV}}
\def \astrobj#1{#1}


\title{Planetary nebulae that cannot be explained by binary systems}

\author{Ealeal Bear\altaffilmark{1} and  Noam Soker\altaffilmark{1}}

\altaffiltext{1}{Department of Physics, Technion -- Israel
Institute of Technology, Haifa 32000, Israel; ealeal@physics.technion.ac.il,
soker@physics.technion.ac.il}

\begin{abstract}
We examine the images of hundreds of planetary nebulae (PNe) and find that for about one in six PNe the morphology is too `messy' to be accounted for by models of stellar binary interaction. We speculate that interacting triple stellar systems shaped these PNe. In this preliminary study we qualitatively classify PNe by one of four categories. (1) PNe that show no need for a tertiary star to account for their morphology. (2) PNe whose structure possesses a pronounced departure from axial-symmetry and/or mirror-symmetry. We classify these, according to our speculation, as `having a triple stellar progenitor'. (3) PNe whose morphology possesses departure from axial-symmetry and/or mirror-symmetry, but not as pronounced as in the previous class, and are classified as `likely shaped by triple stellar system'. (4) PNe with minor departure from axial-symmetry and/or mirror symmetry that could have been as well caused by an eccentric binary system or the inter-stellar medium. These are classified as `maybe shaped by a triple stellar system'.
Given a weight $\eta_t=1, \eta_l=0.67, \eta_m=0.33$ to classes 2, 3 and 4, respectively, we find that according to our assumption about $13 - 21\%$ of PNe have been shaped by triple stellar systems.
Although in some evolutionary scenarios not all three stars survive the evolution, we encourage the search for a triple stellar systems at the center of some PNe.
\end{abstract}


\section{INTRODUCTION}
\label{sec:intro}

 The mass-loss rate and morphology from evolved stars are sensitive to the presence of close stellar and sub-stellar companions. Planetary nebulae (PNe) form the richest group of nebulae around evolve stars. The study of central binary systems of PNe alongside their morphologies, e.g., the alignment of the binary axis with the PN symmetry axis  \citep{Hillwigetal2016b} can teach us a lot about the strong interaction of binary stellar systems (e.g, \citealt{DeMarco2015, Zijlstra2015}).
The majority of studies have concentrated on binary system progenitors of PNe and proto-PNe, which either experienced a common envelope evolution (CEE; e.g., limiting the list to papers from 2016,
\citealt{Chiotellisetal2016, DeMarcoetal2016, GarciaRojasetal2016, Harveyetal2016, Heoetal2016, Hillwigetal2016a, Hillwigetal2016b, Jones2016, Jonesetal2016}), or avoided a CEE (e.g., \citealt{Lagadecetal2011, Decinetal2015, Gorlovaetal2015}). Recently the interest in triple-stellar evolution with mass transfer and mass-loss has grown (e.g., \citealt{MichaelyPerets2014, Portegies2016}), including in PNe (e.g., \citealt{Jones2016}).

Only a small number of papers considered the shaping of PNe by triple stellar systems. These studies include the role of a wide tertiary star at orbital separations of $a_3 \approx 10-{\rm several}\times 10^3 \AU$ in causing a departure from axisymmetry, such as in the PN NGC~3242 \citep{Sokeretal1992}, and the formation of an equatorial spiral pattern \citep{Soker1994}.
Another setting is of the tight binary system that accretes mass from the wind of the AGB star, and one or two of the stars launch jets. When the inner orbital plane of the tight binary system is inclined to the orbital plane of the tight binary system around the AGB star, the morphology of the descendant PN might be lacking any type of symmetry \citep{Soker2004}, i.e., be a `messy PN' (highly irregular), as demonstrated by \cite{AkashiSoker2017}.
\cite{Soker2004} listed several PNe that might have been shaped by a triple stellar progenitor, including IC~2149, NGC~6210, and NGC~1514 that we also mention in the present study.

A tertiary star might influence the mass-loss process as to impose a departure from any kind of symmetry in the descendant PN. Although binary systems, such as those of eccentric orbits, can also cause departure from axisymmetry (e.g.,  \citealt{SokerHadar2002}), they cannot lead to a departure from mirror-symmetry and point-symmetry together.
Very wide tertiary stars, or even higher order hierarchical systems at hundreds of AU (e.g., as found by \citealt{AdamMugrauer2014} in the PN NGC~246), cannot play a role in shaping the morphologies studied in the present study.

An interesting case is the PN SuWt~2. \cite{Exteretal2010} propose that a tight binary system composed of A-stars were engulfed by the AGB stellar progenitor of the PN SuWt~2, and survived the CEE. They further suggest that triple-stellar interaction might eject a high density equatorial ring (see also \citealt{Bondetal2002}). However, \cite{JonesBoffin2017} conclude that the binary A-type stars is a field star system, by chance lying in the same line of sight as the nebular center, and it has no relation to the PN SuWt~2. Our morphological classification of SuWt~2 is compatible with the finding of \cite{JonesBoffin2017}.

In an earlier exploratory study \cite{Soker2016triple} lists several possible processes by which a tight binary system can influence the mass-loss geometry from an AGB star, and form a `messy' PN, including a progenitor AGB star that swallowed a tight binary system.
The tight binary system can survive the CEE \citep{Exteretal2010}, or else it is destroyed by merger with the AGB core or by tidal breaking-up inside the AGB envelope. One of the stars of the broken-up tight binary system might leave the system, or alternatively lose angular momentum in the break-up process and `be drowning' in the AGB envelope, even to the point of merging with the AGB core. As well, one star might leave the system while the other star is drowning.  In an alternative destruction process of the tight binary system, the two stars merge with each other and release a large amount of gravitational energy, leading to a very asymmetrical envelope ejection. A `messy' PN is born. \cite{Soker2016triple} assumes that messy PNe are descendant of triple stellar evolution, and estimates that about one in eight non-spherical PNe have evolved through one of these triple-stellar evolutionary routes.

In this still preliminary study, we set two goals. The first is to list many PNe that we speculate, based on a qualitative analysis of their morphology, were likely shaped by a triple stellar evolution. In some of them the three stars might have survived, and a search for a tertiary star is encouraged. The second goal is to examine the fraction of PNe that were likely shaped by triple-stellar evolution.
We sort the PNe to five classes that we describe in section \ref{sec:classes}. In section \ref{sec:numbers} we list the number of PNe in each class, and estimate the fraction of PNe that are shaped by a triple stellar progenitor. Our short summary is in section \ref{sec:summary}.

\section{MORPHOLOGY CLASSES}
 \label{sec:classes}

We distinguish only between classes according to the departure from axisymmetry, and do not consider the axisymmetrical morphology.

(1) \emph{Not shaped by triple-stellar system.} These are PNe with well resolved images that have nice axisymmetrical (including spherical) or point-symmetric morphologies, and there is no indication for any influence of a tertiary star (although a tertiary star at a large separation or a sub-stellar tertiary body might be present). These PNe are usually classified  as bipolar, elliptical and spherical. There are many examples for such PNe, e.g.,
PN~G215.2-24.2 [IC 418] (NASA and The Hubble Heritage Team with acknowledgment to Raghvendra Sahai and Arsen R. Hajian) and PN~G349.5+01.0 [NGC 6302] \citep{Schwarzetal1992}. Further classification of the different types of axisymmetrical morphologies can be found in, e.g., \cite{Sahaietal2011} and \cite{Stanghellinietal2016}.

\begin{figure}[!t]
\centering
\includegraphics[trim= 0.0cm 0.0cm 0.0cm 0.0cm,clip=true,width=0.6\textwidth]{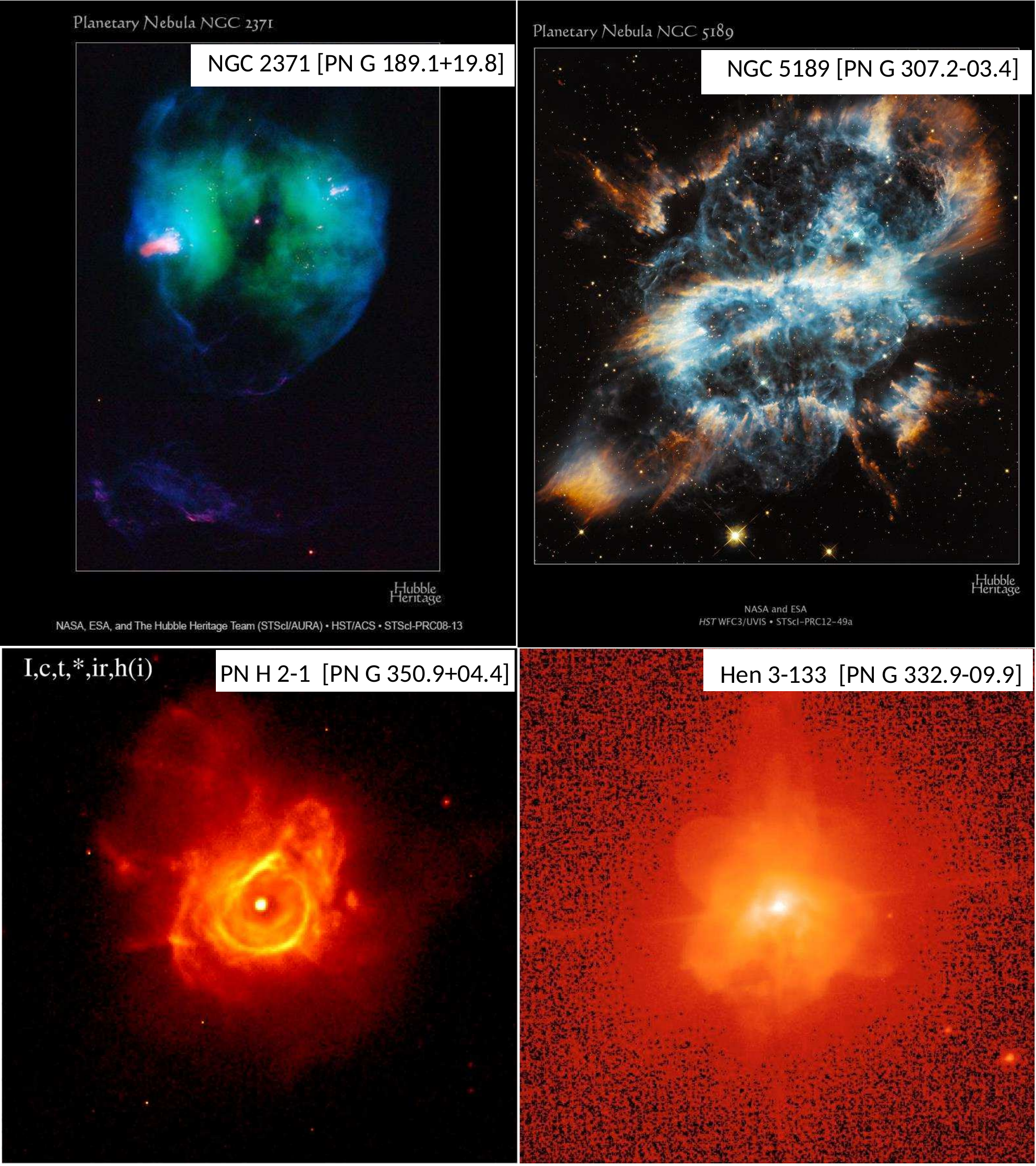}
\caption{Examples of four PNe to which we attribute high probability that they were shaped by a triple-stellar system (the \emph {triple progenitor} class) .
These PNe share large deviations from any structure of symmetry. They contain neither a symmetry plane, nor a symmetry axis, nor a point-symmetric morphology. More examples are presented by \cite{Soker2016triple}. The PN NGC~5189 has a central binary system with an orbital period of 4.05~d \citep{Manicketal2015}. Credits are as follows.
NGC~2371 [PN~G189.1+19.8]: NASA, ESA, and Hubble Heritage Team (STScI/AURA); see also \cite{RamosLariosPhillips2012}. NGC~5189 [PN~G307.2-03.4]: NASA, ESA, and Hubble Heritage Team (STScI/AURA); see also \cite{Sabinetal2012}.
H~2-1 [PN~G350.9+04.4]: \cite{Sahaietal2011}.
Hen~3-1333 [PN~G332.9-09.9; CPD-56◦8032]: \cite{Chesneauetal2006}; see also \cite{DanehkarParker2015} and \cite{Sahaietal2011}.}
  \label{fig:class_Triple}
\end{figure}

(2) \emph{Shaped by a triple stellar progenitor.} The morphologies of PNe in this class possess large departures from axisymmetrical and point-symmetrical structures, and in some cases from mirror-symmetry as well. We do not see signatures of interaction with the ISM, and assume that three stars, the AGB progenitor and two other stars, influenced the outflow and caused these departures from axisymmetry. Some possible triple-stellar evolutionary routes that lead to such features are discussed by \cite{Soker2016triple}.
We present images of four such PNe in Fig. \ref{fig:class_Triple}.

\begin{figure}[!t]
\centering
\includegraphics[trim= 0.0cm 0.0cm 0.0cm 0.0cm,clip=true,width=0.8\textwidth]{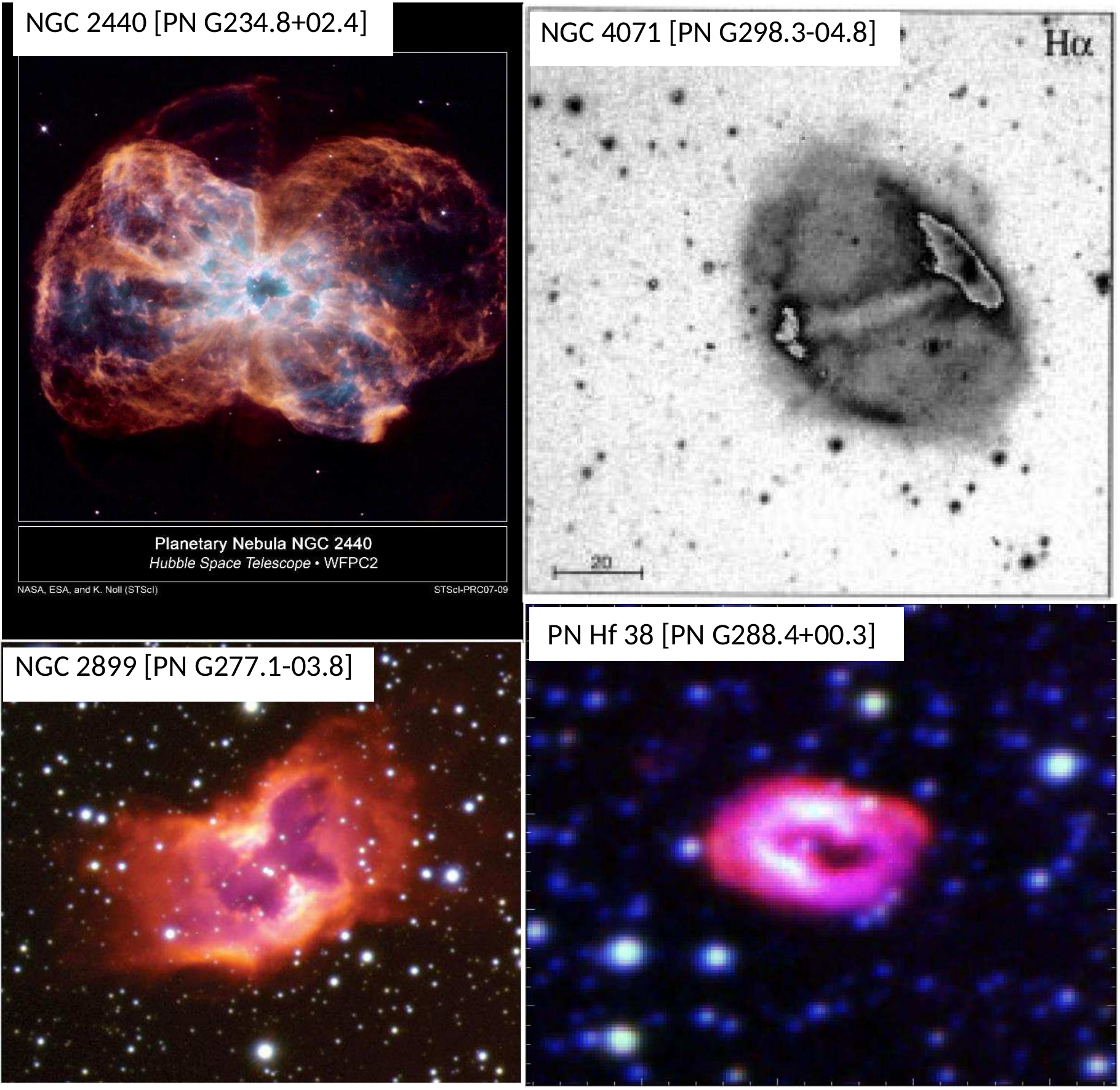}
\caption{Examples of four PNe to which we attribute medium probability that they were shaped by a triple-stellar system (the \emph{likely class}).
These PNe share deviations from any structure with a symmetry. They contain neither a symmetry plane, nor a symmetry axis, nor a point-symmetry. These deviations are smaller than those in the PNe presented in Fig. \ref{fig:class_Triple}. Possible alternative explanations for the deviations from axisymmetry might be as follows. For NGC 2440 an interaction with the ISM (\citealt{RamosLariosPhillips2009}; recently \citealt{LagoCosta2016} found that a torus that is segmented at least in three parts, reproduce the observations). For NGC 4071 an interaction with a dense ISM {{{{ might explain the unequal bright regions on the two sides, but it cannot explain the tilted
"equatorial" stripe. }}}} For NGC 2899 bending of jets launched by the secondary star; they are bent by a wind from the primary star. For PN~G288.4+00.3 an alternative explanation might be an interaction with a very dense ISM cloud or entering a CEE of an eccentric binary system.
Credit for images are as follows. NGC2440 [PN~G234.8+02.4]:  NASA, ESA, and K. Noll (STScI).
NGC 4071 [PN~G298.3-04.8]: \cite{Gornyetal1999}.
NGC2899 [PN~G277.1-03.8]: \cite{Drewetal2014}.
Hf~38 [PN~G288.4+00.3]: \cite{Parkeretal2016} and \cite{Bojicicetal2016} .}
 \label{fig:class_Likely}
\end{figure}

(3) \emph{Likely shaped by a triple stellar progenitor.} These PNe possess departures from axisymmetrical and point-symmetrical structures, and in some cases from mirror-symmetry as well, but not as pronounced as in the \emph {triple progenitor} (class 2) presented in Fig. \ref{fig:class_Triple}. We present images of four such PNe in Fig. \ref{fig:class_Likely}. Many of them were shaped by a triple-stellar system, but other explanations are possible for this class.
For example, He(n)~2-428 [PN~G049.4+02.4 in Table \ref{tab:PNICHASH_PNe}] is classified in this paper as likely triple. However, the alternative might be ISM interaction as suggested by \cite{SokerHadar2002} who classified this PN as `Typical bipolar PN' with `Departure in the outer region'.
Some possible alternative explanations for the departures of the morphologies from axisymmetry are listed in the caption of Fig. \ref{fig:class_Likely}.

\begin{figure}[!t]
\centering
\includegraphics[trim= 0.0cm 0.0cm 0.0cm 0.0cm,clip=true,width=0.8\textwidth]{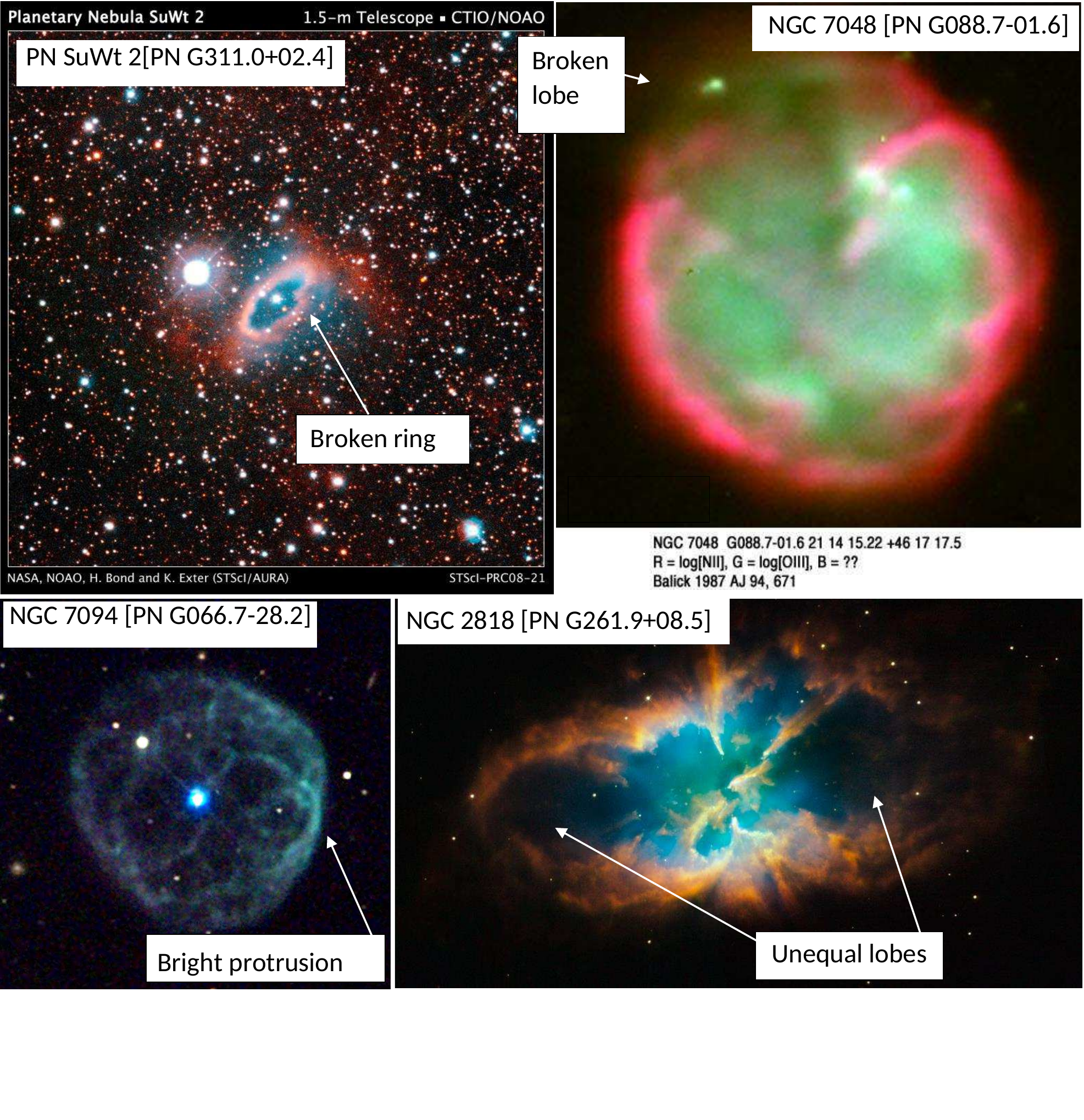}
\caption{Like Fig. \ref{fig:class_Likely}, but for PNe to which we attribute low probability that they were shaped by a triple-stellar system (the \emph{maybe class}).
For these PNe we consider the alternative explanations to be more likely than a triple stellar shaping.
The broken ring in SuWt~2 could be the result of the onset of a CEE, the unequal lobes of NGC 7048 and NGC 2818 could result from instabilities in the process of launching the jets, and NGC 7094 could have been shaped by the ISM.
Credit for images are as follows. SuWt~2 [PN~G311.0+02.4]: \cite{Exteretal2010}.
NGC 7048 [PN~G088.7-01.6]: \cite{Balick1987}.
NGC 7094 [PN~G066.7-28.2]: Adam Block/NOAO/AURA/NSF.
NGC 2818 [PN~G261.9+08.5]: NASA, ESA, and Hubble Heritage Team (STScI/AURA); see also \cite{Vazquez2012}.}
  \label{fig:class_Maybe}
\end{figure}

(4) \emph{Maybe shaped by a triple stellar progenitor.}  These PNe, the \emph{maybe class}, possess small departures from axisymmetrical and point-symmetrical structures. It is less secure to conclude that they have been shaped by a triple stellar system. For example, IC 4997 [PN~G058.3-10.9 in Table \ref{tab:PNICHASH_PNe}] is classified by \cite{SokerHadar2002} as `typical bipolar PN' and `bent'. This bending was attributed to a wide binary companion which lost mass in the AGB phase (for more details see \citealt{SokerHadar2002}).
 The possible alternative explanations for the departures of the morphologies of these PNe from axisymmetry are listed in the caption of Fig. \ref{fig:class_Maybe}.

(5) \emph{Cannot tell.} For PNe belonging to this group we could not tell whether a shaping by a triple stellar system might have taken place. There are two main types of images that fall into this classification.
In the first type the inner region is not resolved, e.g., as for the PN~G118.0-08.6 [Vy 1-1] \citep{Manchadoetal1996,Parkeretal2016}. In the second type the PN seems to have been significantly shaped by interaction with the inter-stellar medium (ISM), e.g., there is a bow-shape structure in the outer regions as in PN~G219.1+31.2 [Abell 31] (SkyHound catalog) and PN~G204.0-08.5 [A 13], \citep{Manchadoetal1996,Parkeretal2016}. Moreover, the PN might be located in a dense stellar population that hints on a dense ISM influencing the shaping, such as for PN~G192.5+07.2 [HDW 6] \citep{Manchadoetal1996} and for PN~G336.8-07.2 [K 2-17] \citep{Schwarzetal1992,Parkeretal2016}.

The above is a qualitative and somewhat subjective scheme of classification based on visual inspection of images of PNe. This  qualitative-subjective classification might be  wrong for some particular PNe. Indeed, there were PNe that the two authors attributed to different classes. However, by the two authors examining independently tens of images, we found that the general statistics holds the same despite some particular PNe that are classified differently. For that, we hold that the statistical conclusion, that is the main goal of the present study, is robust.

\section{STATISTICS}
 \label{sec:numbers}
We have reviewed 656 PNe from the Planetary Nebula Image Catalogue (PNIC), compiled by Bruce Balick (started with his seminar paper \citealt{Balick1987}). We also examined  {3494 PNe} from the HASH catalog \citep{Parkeretal2016, Bojicicetal2016}, most of which belong to the class of `can not tell'.  {{{{ The PNe in the HASH catalog are classified as to the confidence in being a true PN (e.g., 'True PN', 'Likely PN', 'Possible PN'). We only include in our statistics (see Eq. \ref{triples2}) `True PNe'.
In Table \ref{tab:PNICHASH_PNe} we list all the PNe from the PNIC catalog that belong to one of the categories of not shaped by a binary system (triple stellar system, likely triple, and maybe triple). We also list several PNe from the HASH catalog; we list the rest of the PNe from the HASH catalog that belong to the classes of not shaped by a binary system in Table A1 in the appendix. }}}}
\begin{table}
\caption{PNe from the PNIC and HASH catalogs according to their classification. We list all the PNe from the HASH catalog that were not shaped by a binary system in Table A1 (full table is available in the on-line material of ApJL 2017, 837, L10).
\label{tab:PNICHASH_PNe} }
\begin{tabular}{ccc}
  \hline
  Maybe triple [PNIC]  & Likely triple [PNIC]    & Triple progenitor [PNIC]    \\
    \hline
  PN~G003.7+07.9 [H 2-8]        & PN~G026.6-01.5 [K 4-5]        & PN~G035.9-01.1 [Sh 2-71] \\
  PN~G006.4+02.0 [M 1-31]       & PN~G049.4+02.4 [He 2-428]     & PN~G043.1+37.7 [NGC 6210] \\
  PN~G024.3-03.3 [Pe 1-17]      & PN~G167.4-09.1 [K 3-66]       & PN~G057.2-08.9 [NGC 6879] \\
  PN~G036.9-01.1 [HaTr 11]      &  	PN~G358.9-00.7 [M 1-26/He 2-277]	   & PN~G165.5-15.2 [NGC 1514]\\%
  PN~G045.7-04.5 [NGC 6804]     &  PN~G174.2-14.6 [H 3-29]      & PN~G166.1+10.4 [IC 2149]  \\
  PN~G050.4+05.2 [A 52]         &  PN~G194.2+02.5 [J 900]       & PN~G189.1+19.8 [NGC 2371-2]\\
  PN~G051.4+09.6 [Hu 2-1]       &  PN~G234.8+02.4 [NGC 2440]    & PN~G292.6+01.2 [NGC 3699] \\
  PN~G056.0+02.0 [K 3-35]       &  PN~G259.1+00.9 [He 2-11]     & PN~G296.4-06.9 [He 2-71] \\%
  PN~G057.9-01.5 [He 2-447]     &  PN~G277.1-03.8 [NGC 2899]    & PN~G350.9+04.4 [H 2-1]  \\
  PN~G058.3-10.9 [IC 4997]      &  PN~G285.4-05.3 [IC 2553]     & PN~G307.2-03.4 [NGC 5189] \\ 
  PN~G058.6+06.1 [A 57]         &  PN~G293.6+01.2 [He 2-70]     & PN~G321.0+03.9 [He 2-113] \\
  PN~G059.0+04.6 [K 3-34]       &  PN~G298.3-04.8 [NGC 4071]    & PN~G332.9-09.9 [He 3-1333] \\
  PN~G059.7-18.7 [A 72]         &  PN~G315.0-00.3 [He 2-111]    & PN~G342.9-04.9 [He 2-207] \\
  PN~G060.0-04.3 [A 68]         &  PN~G341.2-24.6 [Lo 18]       & PN~G345.4+00.1 [IC 4637] \\
  PN~G066.7-28.2 [NGC 7094]     &  PN~G341.8+05.4 [NGC 6153]    &   \\
  PN~G088.7-01.6 [NGC 7048]     &  PN~G342.5-14.3 [Sp 3]        &   \\
  PN~G201.9-04.6 [We 1-4]       &  PN~G355.4-04.0 [Hf 2-1]     &   \\
  PN~G205.1+14.2 [A 21]         &        &   \\
  PN~G242.6-11.6 [M 3-1]        &                                &   \\
  PN~G261.9+08.5 [NGC 2818]     &                               &   \\
  PN~G283.9-01.8 [Hf 4]         &                               &   \\
  PN~G311.0+02.4 [SuWt 2]       &                               &   \\
  PN~G312.3+10.5 [NGC 5307]     &                               &   \\
  PN~G317.1-05.7 [NGC 5844]     &                               &   \\
  PN~G318.4+41.4 [A 36]         &                               &   \\
  PN~G322.5-05.2 [NGC 5979]     &                               &   \\
  PN~G325.4-04.0 [He 2-141]     &                               &   \\
  PN~G331.5-02.7 [He 2-161]     &                               &   \\
  PN~G338.1-08.3 [NGC 6326]     &                               &   \\
  PN~G357.0+02.4 [M 4-4]        &                               &   \\
  \hline
    Maybe triple [HASH]  & Likely triple [HASH]    & Triple progenitor [HASH]    \\
    \hline
  PN~G126.6+01.3 				& PN~G229.6-02.7 				& PN~G165.5-15.2 \\
PN~G129.2-02.0 					& PN~G234.8+02.4 				& PN~G166.1+10.4 \\
PN~G131.5+02.6 					& PN~G231.4+04.3 				& PN~G222.9-01.1 \\
PN~G138.1+04.1 					& PN~G277.1-03.8 				& PN~G231.8+04.1 \\
PN~G139.3+04.8 					& PN~G277.7-03.5 				& PN~G235.7+07.1 \\
 \hline
\end{tabular}
\end{table}

Most of the PNe in the HASH catalog contain an overexposed internal region that prevents us from classifying these PNe. For that we will use the PNIC as a minimum estimate and the HASH as maximum estimate for the effect of tertiary star on the shaping of PNe. 
 We note that alternative explanations to the distorted morphology include eccentric binary (e.g., \citealt{SokerHadar2002}), ISM interaction (e.g., \citealt{TweedyKwitter1996, Rauchetal2000}), wide binary companions (e.g., \citealt{Soker1994}) and precession of a jet launched by a binary companion. In the later case, one jet might be tilted towards the denser AGB wind, hence the two jet-inflated lobes on the two sides of the equatorial plane will not be equal. The numbers of PNe we attributed to each of the five classes described in section \ref{sec:classes}, and for the two catalogs, are listed in Table \ref{tab:PNICHASH_Summary}.
 \begin{table}
\caption{The number of PNe in each class from the PNIC (compiled by Bruce Balick) and HASH catalogs (\citealt{Parkeretal2016, Bojicicetal2016}; updated as of June 2nd 2016). {Lines 2 - 5 include only `True PNe' according to the HASH catalog}.
\label{tab:PNICHASH_Summary}}
\centering
\begin{tabular}{lcc}
  \hline
  Class & PNIC & HASH\\
  \hline
  Cannot tell &                                 381 & 3122 \\
  Not shaped by triple-stellar system &         214 & 221 \\
  Shaped by a triple stellar progenitor &       14 & 27\\
  Likely shaped by a triple stellar system &    17 & 31\\
  Maybe shaped by a triple stellar system &     30 & 93\\
  Total &                                       656& 3494\\ 
  Total for statistics&                         275& 372\\
  \hline
\end{tabular}
\end{table}

The study of the PN  M2-29 [PN~G004.0-03.0] demonstrates the `tension' between a triple stellar system progenitor and an eccentric binary.  \cite{Hajduketal2008} suggested that this PN was shaped by a triple stellar system. They claimed that this system includes two stars that orbit the central star of this PN. The close stellar companion has an orbital period of about 23 days, and the tertiary star orbits the inner binary system with an orbital period of about 18 years. In a follow-up paper \citep{Gesickietal2010} they abandoned the triple stellar explanation, and instead considered a binary system in an eccentric orbit to account for the azimuthal asymmetry. The orbital period of the binary system in their new model is in the general range of 200 days to about 100 years. The PN M2-29 is not listed in the PNIC and in the HASH it is classified by us as cannot tell by the morphology, so it is not in our tables. According to its morphology we would have classified it to the likely shaped by triple stellar system group.

We now crudely estimate the fraction of PNe that we expect to have been shaped by a triple stellar system, under the assumptions of the classification.
We simply attribute to each PN in each of the four classes where we could tell the morphology, a probability of being shaped by a triple stellar progenitor.
We assign a probability $\eta_n=0$ to each PN in the class of not-shaped by a triple stellar system.
The probabilities of the other classes are assumed simply and very crudely to be
$\eta_m=\frac{1}{3}$ for the \emph{maybe triple} class, $\eta_l=\frac{2}{3}$ to the \emph{likely triple} class, and $\eta_t=1$ for the \emph{triple progenitor} class.
The total number of PNe in the four classes is 275.
We then calculate the probability of being shaped by a triple stellar progenitor according to the classification of the PNIC
\begin{equation}
 \label{triples1}
 P ({\rm PNIC}) \approx \frac{30 \eta_m + 17 \eta_l   + 14 \eta_t}{275} \approx 13\%.
\end{equation}
The same calculation for the classification of PNe from the HASH catalog, where there are total of 372 PNe in the four relevant classes, gives
\begin{equation}
 \label{triples2}
P ({\rm HASH}) \approx \frac{ 93 \eta_m  + 31 \eta_l + 27 \eta_t}{372} \approx 21\%.
\end{equation}
The HASH catalog contains many PNe in the \emph{cannot tell} class. Some of the PNe for which we could not tell whether they have been shaped by a triple stellar system, are resolved in the PNIC.
For these PNe we took the classification according to our classification from the PNIC.%

Over all we find that the probability of a PN to be shaped in a noticeable manner by a triple stellar progenitor is $P({\rm triple}) \approx 13-21 \%$, or about one in six.
\cite{Soker2016triple} estimates that about one in eight ($\approx 12.5 \%$) of non-spherical PNe is shaped by a process by which the AGB progenitor swallows a tight binary system. Considering other triple-stellar evolutionary routes that do not contain a tight binary system, like a tertiary star further out, the fraction is higher.
Therefore, the probability found from our classification, $P({\rm triple}) \approx 17\%$, and that estimated by \cite{Soker2016triple} are compatible.

\section{SUMMARY}
\label{sec:summary}

We have examined hundreds of images of PNe from two catalogs, as summarized in Table \ref{tab:PNICHASH_Summary}. We then apply qualitative criteria for classifying them according to their departure from spherical symmetry, axial symmetry, and point symmetry. We conjecture that these types of departures might result from the interaction of triple-stellar systems \citep{Soker2016triple}. We then attributed different confidence levels to this conjecture (section \ref{sec:classes}).
We consider PNe with large departures from any kind of symmetry, like the four examples presented in Fig. \ref{fig:class_Triple}, to have been shaped by a triple-stellar system. Namely, we take the probability that the progenitor of these PNe were triple stellar system to be almost 100 per cents, $\eta_t = 1$.
We then built two other classes where the confidence levels are lower, $\eta_l \approx 0.67$ for the \emph{likely class} with four examples presented in Fig. \ref{fig:class_Likely}, and $\eta_m \approx 0.33$ for the \emph{maybe class} with four examples presented in Fig. \ref{fig:class_Maybe}.

Our analysis under the above conjectures and the crude confidence levels, suggest that $\approx 13-21 \%$ of all PNe were shaped by interacting triple stellar progenitor (section \ref{sec:numbers}).
Namely, about one in six PNe are shaped by a strongly interacting triple stellar system.
As discussed by \cite{Soker2016triple}, this is compatible with what we know about the initial mass function of multiple stellar systems.

Although the subjective visual inspection might lead different researchers to classify particular PNe to different classes, the overall statistical result is robust and will not depend on the classifying person (within the range of $\approx 13-21 \%$). For that, this preliminary study has one important guideline for future observations: look for a triple stellar system inside `messy PNe'.

In some cases (beyond the topic of this paper), all three stars might survive (e.g., \citealt{Exteretal2010}). Therefore, the search for a tertiary star in known binary central stars of PNe with highly deformed morphologies is encouraged.
It is also possible that some AGB stars that have non-spherical mass-loss geometry interact with two stars. In these cases the search for triple stellar system is also encouraged.
In some cases the tertiary object might be sub-stellar, namely a brown dwarf or a massive planet. In the latter, external planets might survive.

{{{{ We thank an anonymous referee for very helpful comments. }}}}
N.S. is supported by the Charles Wolfson Academic Chair.
The Planetary Nebula Image Catalogue (PNIC) compiled by Bruce Balick
 \newline
(http://www.astro.washington.edu/users/balick/PNIC/),
and the HASH catalog \citep{Parkeretal2016}, were essential tools in this study.


\label{lastpage}

\end{document}